\documentclass[useAMS,usenatbib]{mn2e}
\usepackage{graphicx,fleqn,fix2col,rotating}
\DeclareGraphicsExtensions{.eps,.ps,.eps.gz,.ps.gz,.eps.Z}
\title[Radial velocity measurements of Subdwarf B stars]{Radial velocity measurements of Subdwarf B stars}

\author[C.M.~Copperwheat et al.]{C.M.~Copperwheat$^{1}$, L.~Morales-Rueda$^{2}$, T.R.~Marsh$^{1}$, \newauthor P.F.L.~Maxted$^{3}$ and U.~Heber$^{4}$\\\\
$^{1}$ Department of Physics, University of Warwick, Coventry, CV4 7AL, UK\\
$^{2}$ Symetrica Security Ltd., Phi House, Southampton Science Park, Southampton, SO16 7NS, UK\\
$^{3}$ Astrophysics Group, Keele University, Keele, Staffordshire, ST5 5BG, UK\\
$^{4}$ Dr. Remeis-Sternwarte, Astronomisches Institut der Universit\"at Erlangen-N\"urnberg, Sternwartstr. 7, 96049 Bamberg, Germany\\
}

\date{Received: }

\begin{document}

\newcommand{\dg} {^{\circ}}
\outer\def\gtae {$\buildrel {\lower3pt\hbox{$>$}} \over
{\lower2pt\hbox{$\sim$}} $}
\outer\def\ltae {$\buildrel {\lower3pt\hbox{$<$}} \over
{\lower2pt\hbox{$\sim$}} $}
\newcommand{\ergscm} {erg s$^{-1}$ cm$^{-2}$}
\newcommand{\ergss} {erg s$^{-1}$}
\newcommand{\ergsd} {erg s$^{-1}$ $d^{2}_{100}$}
\newcommand{\pcmsq} {cm$^{-2}$}
\newcommand{\ros} {{\it ROSAT}}
\newcommand{\xmm} {\mbox{{\it XMM-Newton}}}
\newcommand{\exo} {{\it EXOSAT}}
\newcommand{\sax} {{\it BeppoSAX}}
\newcommand{\chandra} {{\it Chandra}}
\newcommand{\hst} {{\it HST}}
\newcommand{\subaru} {{\it Subaru}}
\def\rchi{{${\chi}_{\nu}^{2}$}}
\newcommand{\Msun} {$M_{\odot}$}
\newcommand{\Mwd} {$M_{wd}$}
\newcommand{\Mbh} {$M_{\bullet}$}
\newcommand{\Lsun} {$L_{\odot}$}
\newcommand{\Rsun} {$R_{\odot}$}
\newcommand{\Zsun} {$Z_{\odot}$}
\def\Mdot{\hbox{$\dot M$}}
\def\mdot{\hbox{$\dot m$}}
\def\mincir{\raise -2.truept\hbox{\rlap{\hbox{$\sim$}}\raise5.truept
\hbox{$<$}\ }}
\def\magcir{\raise -4.truept\hbox{\rlap{\hbox{$\sim$}}\raise5.truept
\hbox{$>$}\ }}
\newcommand{\mnras} {MNRAS}
\newcommand{\aap} {A\&A}
\newcommand{\apj} {ApJ}
\newcommand{\apjl} {ApJL}
\newcommand{\apjs} {ApJS}
\newcommand{\aj} {AJ}
\newcommand{\pasp} {PASP}
\newcommand{\aaps} {AAPS}
\newcommand{\apss} {Ap\&SS}
\newcommand{\araa} {ARAA}
\newcommand{\nat} {Nature}
\newcommand{\pasj} {PASJ}
\newcommand{\ha}{\hbox{$\hbox{H}\alpha$}}
\newcommand{\hb}{\hbox{$\hbox{H}\beta$}}
\newcommand{\hg}{\hbox{$\hbox{H}\gamma$}}
\newcommand{\heii}{\hbox{$\hbox{He\,{\sc ii}\,$\lambda$4686\,\AA}$}}
\newcommand{\hei}{\hbox{$\hbox{He\,{\sc i}\,$\lambda$4472\,\AA}$}}

\maketitle

\begin{abstract} 
Subdwarf B (sdB) stars are hot, sub-luminous stars which are thought to be core-helium burning with thin hydrogen envelopes. The mechanism by which these
stars lose their envelopes has been controversial but it has been argued that binary star interaction is the main cause. Over the past decade we have conducted a radial velocity study of a large sample of sdB stars, and have shown that a significant fraction of the field sdB population exists in binary systems. In 2002 and 2003 we published $23$ new binary sdB stars and the definitions of their orbits. Here we present the continuation of this project. We give the binary parameters for $28$ systems, 18 of which are new. We present also our radial velocity measurements of a further $108$ sdBs. Of these, $88$ show no significant evidence of orbital motion. The remaining $20$ do show radial velocity variations, and so are good candidates for further study. Based on these results, our best estimate for the binary fraction in the sdB population is $46$ -- $56$ per cent. This is a lower bound since the radial velocity variations of very long period systems would be difficult to detect over the baseline of our programme, and for some sources we have only a small number of measurements.
\end{abstract}

\section{INTRODUCTION}  

Subdwarf B stars (sdBs) are hot (T$_{\rm eff}= 25$,$000$ -- $40$,$000\,{\rm K}$), core helium-burning stars with masses $\sim$$0.5$\Msun \ and thin hydrogen envelopes of mass $\le 0.02$\Msun \ (\citealt{Heber84,Saffer94}, see also \citealt{Heber09} for a recent review). \citet{DCruz96} proposed that such a star could form if a red giant star with a degenerate helium core were to lose its envelope when it is within $\sim$$0.4$ mag of the tip of the red giant branch (RGB). In this scenario, the core could go on to ignite helium despite the mass loss. This model explains the masses of sdB stars as a consequence of the core mass at the helium flash. 

The loss of the hydrogen envelope could be due to an enhancement of the stellar wind, or binary interactions. It has been shown that a large fraction of sdB stars are now members of short period binary systems \citep{Maxted01,Napiwotzki04}. Close binary systems such as these imply a `common envelope' phase, which occurs due to dynamically unstable transfer when the more massive star reaches the red giant phase of its evolution. Orbital energy is then lost to friction, resulting in a shortening of the binary period \citep{Iben93}. The common envelope is eventually ejected, leading to an sdB primary star with a close main sequence or white dwarf companion \citep{Han02,Han03}. \citet{Yungelson05} argued that all sdB stars come from binaries, with the apparently single examples either the product of merging pairs of helium white dwarfs or members of long-period systems which have avoided the common envelope phase \citep{Green01}, and whose radial velocity variations we have yet to detect.

SdB stars are of interest because they are a strong test of population synthesis models, since they are much less influenced by the selection effects which compromise other close binary populations, such as cataclysmic variables \citep{Pretorius07}. They are also an ideal population for testing models of the common envelope phase, and sdBs with massive white dwarf companions are one of the potential progenitors for type Ia supernovae \citep{Tutukov81, Webbink84}.

Following the detection of many close binary sdBs by \citet{Maxted01}, we began a project with the aim of measuring the orbits of a large number of binary systems. In \citet{Maxted02} and 
\citet{MoralesRueda03} we published the parameters of 23 systems. We present here the continuation of this study.

\section{OBSERVATIONS AND REDUCTION}

\begin{table*}
\caption{Journal of observations.}
\label{tab:obs}
\begin{center}
\begin{tabular}{lcl}
Dates & \# of RV observations & Setup \\
\hline
10 -- 21 Apr 2000       & 18    & INT + IDS + 500 mm + R1200R + $\lambda_c$=\ha \\
8 -- 13 Mar 2001        &72     & INT + IDS + 500 mm + R1200R + $\lambda_c$=\ha \\
1 -- 8 May 2001         & 104    & INT + IDS + 500 mm + R1200R + $\lambda_c$=\ha \\
6 -- 11 Aug 2001        & 76     & INT + IDS + 500 mm + R1200B + $\lambda_c$=4350\AA \\
27 Sept -- 6 Oct 2001   & 135    & INT + IDS + 500 mm + R1200B + $\lambda_c$=4350\AA \\
27 Mar -- 1 Apr 2002    & 175    & INT + IDS + 500 mm + R1200B + $\lambda_c$=4350\AA \\
23 -- 30 Apr 2002       & 217   & INT + IDS + 500 mm + R1200B + $\lambda_c$=4350\AA \\
21 -- 22 Mar 2003       & 35    & SAAO 1.9m + grating spectrograph + grating \#4 + $\lambda_c$=4600\AA \\
9 -- 18 Apr 2003        & 101    & INT + IDS + 500 mm + R1200B + $\lambda_c$=4400\AA \\
10 -- 15 Sep 2003       & 136   & SAAO 1.9m + grating spectrograph + grating \#4 + $\lambda_c$=4600\AA \\
30 Mar -- 7 Apr 2004    & 75    & SAAO 1.9m + grating spectrograph + grating \#4 + $\lambda_c$=4600\AA \\
22 -- 25 Oct 2004       & 107    & SAAO 1.9m + grating spectrograph + grating \#4 + $\lambda_c$=4600\AA \\
22 -- 27 June 2005      & 101    & SAAO 1.9m + grating spectrograph + grating \#4 + $\lambda_c$=4600\AA \\
11 -- 24 Oct 2005       & 223    & SAAO 1.9m + grating spectrograph + grating \#4 + $\lambda_c$=4600\AA \\
25 Mar -- 6 Apr 2007    & 243   & INT + IDS + 500 mm + R1200B + $\lambda_c$=4500\AA \\
17 -- 25 Aug 2007       & 150    & INT + IDS + 500 mm + R1200B + $\lambda_c$=4500\AA \\
18 -- 27 Mar 2008       & 61    & INT + IDS + 500 mm + R1200B + $\lambda_c$=4500\AA \\
10 -- 16 Mar 2009       & 82    & INT + IDS + 500 mm + R1200B + $\lambda_c$=4500\AA \\
30 Apr -- 1 May 2009    & 12     & WHT + ISIS + R600B ($\lambda_c$=4350\AA) + R1200R ($\lambda_c$=\ha)  \\
\end{tabular}
\end{center}
\end{table*}

A complete journal of our observations is given in Table \ref{tab:obs}. The number of radial velocity observations which we list in this table are those which are presented in this paper, in addition to the measurements previously presented in \citet{Maxted02} and \citet{MoralesRueda03}.

The majority of the data used in this study were collected using the Intermediate Dispersion Spectrograph (IDS) mounted on the 2.5m Isaac Newton Telescope. Our earliest observations in 2000/2001 used the 500mm camera with the R1200R grating centred on \ha\ and the TEK (1k $\times$ 1k) CCD giving a dispersion of 0.37\,\AA/pix and a resolution of 0.9\,\AA. Subsequent observations used the 235 mm camera with the R1200B grating and the thinned EEV10 (2k $\times$ 4k) CCD, giving a dispersion of 0.48\,\AA/pix and a resolution of 1.4\,\AA. The central wavelength used with this grating varied slightly between observing runs, but in all cases was chosen to cover \hb \ and \hg. In 2003 we expanded this project to cover southern targets with the first of a series of observing runs using the grating spectrograph plus the SITe CCD mounted on the 1.9\,m telescope at the Sutherland site of the South African Astronomical Observatory (SAAO). Grating 4, with 1200 grooves per mm, was used to obtain spectra covering \hg \ and \hb \ with a dispersion of 0.5\AA/pix and a resolution of better than 1\AA\ at 4600\AA.

We took two consecutive observations of each object and bracketed them with CuAr frames to calibrate the spectra in wavelength. After debiasing and flatfielding the frames (using Tungsten flatfield frames for the INT/IDS observations and dome flats for the SAAO observations), spectral extraction proceeded according to the optimal algorithm of \citet{Marsh89}. The arcs were extracted using the profile associated with their corresponding target to avoid systematic errors caused by the spectra being tilted.  Uncertainties on every point were propagated through every stage of the data reduction.

\section{RESULTS}

\subsection{Radial velocity measurements}
\label{sec:rv}

To measure the radial velocities we use least squares fitting of a model line profile and follow the same procedure described in \citet{MoralesRueda03}. The model line profile consists of three Gaussians with different widths and depths. For any given sdB, the widths and depths of the Gaussians are optimised and then fixed while their velocity offsets from the rest wavelengths of the lines in question are fitted separately for each spectrum; see \citet{Maxted00b} for further details of this procedure. For the red spectra obtained in Apr 2000 - May 2001 we fit the \ha \ line. For all other spectra the fitting was performed simultaneously for \hb \ and \hg.

In Tables \ref{tab:rv1a} and \ref{tab:rv2a} we list all of our radial velocity measurements. Table \ref{tab:rv1a} contains the measurements for 28 systems which we have found to be binary, and for which we have found the orbital period.  The description of the orbital period determination is given in Section \ref{sec:periods}. Table \ref{tab:rv2a} contains measurements for all of the remaining sdBs covered by our project which we have not previously published, a total of 108 objects. Most of these objects show no evidence for orbital variations, and so are likely to be either single sdBs,  or binary systems in which the mass function is too low for us to detect radial velocity variations (due to a low companion mass, a long binary period, a binary inclination close to zero or a combination of these factors). For some objects the non-detection of variation may be because the number of observations of that object is small. In some other cases, there are clear signs that the sdB is in a binary system, however there are still competing orbital aliases of comparable significance and so our data is insufficient to determine the true orbital period. These are strong candidates for future observation. In this paper we will concentrate on the 28 solved systems given in Table \ref{tab:rv1a}. In Section \ref{sec:binfrac} we discuss the binary fraction of the sdB population and in Section \ref{sec:composite} we investigate the nature of the companion stars. For these sections we consider the complete list of objects in our programme.

\subsection{Determination of orbital periods}
\label{sec:periods}

\begin{table*}
\caption{List of the orbital periods measured for the 28 sdBs studied. We give T$_{0}$, the systemic velocity, $\gamma$, the radial velocity
semi-amplitude, K, and the reduced $\chi^2$ achieved for the best alias. We also list the period of the $N$th best alias and the $\chi^2$ difference between the primary and $N$th
aliases. For the majority of targets we list the second best alias, however in the few cases where there are many aliases close in period to the primary alias, we give the first alias for which the period differs from the primary alias by $\pm 5$ per cent. The number of data points used to calculate the orbital period is given in the final column under n. }
\label{tab:results}
\begin{center}
\begin{tabular}{lllr@{\,$\pm$\,}lr@{\,$\pm$\,}llr|lr|r}
Object & HMJD (T$_{0}$) & Period (d)& \multicolumn{2}{c}{$\gamma$ (km/s)} & \multicolumn{2}{c}{K (km/s)} &
 $\chi^{2}_{reduced}$ &$N$ &$N$th best alias (d)& $\Delta \chi^{2}$ & n \\
\hline
EC00404-4429    &52894.9418(4)  &0.12834(4)     &$33.0$&$2.9$       &$152.8$&$3.4$      &0.82 
    &2&0.11350(3)     &74     &9\\
EC02200-2338    &52895.529(4)   &0.8022(7)      &$20.7$&$2.3$       &$96.4$&$1.4$       &0.27
    &2&0.3038(1)      &61     &10\\
PG0919+273      &53274.29(4)    &15.5830(5)     &$-68.6$&$0.6$      &$41.5$&$0.8$       &1.23
    &2&14.9400(5)     &43     &47\\
PG0934+186      &52376.08(1)    &4.05(1)        &$7.7$&$3.2$        &$60.3$&$2.4$       &0.74
    &2&3.59(1)        &22     &18\\
PG0958-073           &53635.03(2)    &3.18095(7)     &$90.5$&$0.8$       &$27.6$&$1.4$       &2.31
    &7&2.92145(5)     &29      &21\\
PG1000+408      &52920.208(7)   &1.049343(5)    &$56.6$&$3.4$       &$63.5$&$3.0$       &0.73
    &123&0.705291(2)    &55     &20\\
PG1230+052      &53276.431(3)   &0.837177(3)    &$-43.1$&$0.7$      &$40.4$&$1.2$       &1.28
    &70&0.737931(2)    &229     &29\\
EC12408-1427    &52954.467(5)   &0.90243(1)     &$-52.2$&$1.2$      &$58.6$&$1.5$       &0.72 
    &2&9.493(1)       &38     &29\\
PG1244+113      &53301.61(2)    &5.75211(9)     &$7.4$&$0.8$        &$54.4$&$1.4$       &1.94
    &2&5.8456(1)      &206    &51\\
PG1253+284      &53673.77(1)    &3.01634(5)     &$17.8$&$0.6$       &$24.8$&$0.9$       &1.83 
    &2&2.95692(6)     &28     &32\\
EC13332-1424    &52954.418(3)   &0.82794(1)     &$-53.2$&$1.8$      &$104.1$&$3.0$      &0.77
    &2&0.455168(3)    &33     &22\\
PG1403+316      &53293.779(6)   &1.73846(1)     &$-2.1$&$0.9$       &$58.5$&$1.8$       &0.88
    &28&1.63874(1)     &34     &16\\
PG1439-013      &53629.87(7)    &7.2914(5)      &$-53.7$&$1.6$      &$50.7$&$1.5$       &1.52
    &2&7.1452(5)      &16     &38\\
PG1452+198      &52262.240(4)   &0.96498(4)     &$-9.1$&$2.1$       &$86.8$&$1.9$       &0.99
    &2&0.97213(4)     &2      &20\\
PG1519+640      &52391.389(3)   &0.539(3)       &$0.9$&$0.8$        &$36.7$&$1.2$       &1.29
    &2&0.338(2)       &68     &18\\
PG1528+104      &52394.553(2)   &0.331(1)       &$-49.3$&$1.0$      &$53.3$&$1.6$       &1.03
    &2&0.491(2)       &33     &14\\
PG1558-007      &53280.25(4)    &10.3495(6)     &$-71.9$&$0.7$      &$42.8$&$0.8$       &1.77
    &2&1.103666(8)    &102    &42\\
PG1648+536      &53650.8209(9)  &0.6109107(4)   &$-69.9$&$0.9$      &$109.0$&$1.3$      &1.63
    &37&0.5500470(5)   &202      &61\\
KUV16256+4034   &52392.498(2)   &0.4776(8)      &$-90.9$&$0.9$      &$38.7$&$1.2$       &1.68
    &2&0.958(2)       &64     &18\\
EC20182-6534    &53276.834(3)   &0.598819(6)    &$13.5$&$1.9$       &$59.7$&$3.2$       &0.84
    &2&0.585875(6)    &40     &25\\
EC20260-4757    & 53279.37(6)   &8.952(2)       &$56.5$&$1.6$       &$57.1$&$1.9$       &2.25
    &2&0.88375(2)     &22     &29\\
EC20369-1804    &53279.29(3)    &4.5095(4)      &$7.2$&$1.6$        &$51.5$&$2.3$       &0.57
    &2&0.81668(1)     &20     &24\\
KPD2040+3955    &53259.232(3)   &1.482860(4)    &$-16.4$&$1.0$      &$94.0$&$1.5$       &3.21
    &2&1.494224(4)    &37     &20\\
EC21556-5552    &53660.604(5)   &0.8340(7)      &$31.4$&$2.0$       &$65.0$&$3.4$       &1.14
    &2&0.4545(2)      &52     &18\\
KPD2215+5037    &53258.172(2)   &0.809146(2)    &$-7.2$&$1.0$       &$86.0$&$1.5$       &2.04
    &147&0.768551(1)    &103     &12\\
EC22202-1834    &53605.762(4)   &0.70471(5)     &$-5.5$&$3.9$       &$118.6$&$5.8$      &0.77
    &2&0.70884(5)     &5      &14\\
EC22590-4819    &53278.25(6)    &10.359(2)      &$13.5$&$1.1$       &$46.8$&$1.8$       &0.84
    &2&0.90970(2)     &46     &26\\
PG2331+038      &53234.865(3)   &1.204964(3)    &$-9.5 $&$ 1.1$     &$93.5 $&$ 1.9$     &2.68 
    &51&0.020232962(1) &282     &18\\
\end{tabular}
\end{center}
\end{table*}

In radial velocity work, while one can very soon determine a star to be binary, it can take much longer to determine the orbital period. With a small amount of data there is a danger of picking an incorrect alias. This period can be very wrong, even when the quoted uncertainty is tiny, because the statistics are not Gaussian and so an error of 100 or even 1000 times the quoted uncertainty can happen. On the other hand, the process of collecting sufficient data to determine the true period beyond doubt can be very inefficient in terms of telescope time. It was therefore necessary for us to select a criterion by which we consider a binary orbit to be solved.

Our initial `rule of thumb' was to consider the period in a system to be determined if the lowest minimum of the $\chi^2$ function has a $\chi^2$ at least 10 less than the next lowest minimum. From a Bayesian point of view, this $\Delta\chi^2>10$ criterion is equivalent to requiring that the second best period is $\ga \exp 5 = 150$ times less probable than the best. This argument, although true, is not precise because while the peak of the second alias may be $> 150$ times less probable than the peak of the best alias, there is no guarantee that the total probability of \emph{any} other period is as low. We therefore instead chose to determine the probabilities of the true orbital period being within a certain range of our favoured value. The details of this Bayesian calculation are given in \citet{MoralesRueda03} and \citet{Marsh95}. For the purposes of comparing the observed sdB binary period distribution to theoretical evolutionary models, knowing the orbital period to within $\pm 5$ per cent is sufficient. We therefore established the following criterion: we considered a system to be solved if the probability that the true orbital period lies further than $5$ per cent from our best alias is less than $0.1$ per cent. 

For 28 of our targets we have enough radial velocity measurements to satisfy this criterion. We have previously announced nine of these in conference proceedings (PG0934+186, PG1230+052, PG1244+113, PG1519+640, PG1528+104, KPD2040+3955, \citealt{MoralesRueda03a}; EC00404-4429, EC02200-2338, \citealt{MoralesRueda05}, EC12408-1427 \citealt{MoralesRueda06}), but we present here the more detailed analysis. The orbit of one of our targets has been independently determined (PG1000+408, \citealt{Shimanskii08}) and so we present our results to corroborate this determination. The remaining 18 orbital determinations are new.

We follow the procedure described in \citet{MoralesRueda03}, using the `floating mean' periodogram (e.g. \citealt{Cumming99}), which consists in fitting the data with a model composed of a sinusoid plus a constant of the form:
\[ V = \gamma + {\rm K}  \sin (2 \pi f (t - t_0)),\]
where $f$ is the frequency ($f$ = $1/$period) and $t$ is the observation time. We obtain the $\chi^2$ of the fits as a function of frequency and select the minima of this $\chi^2$ function. By fitting the systemic velocity, $\gamma$, at the same time as $K$ and $t_0$, we correct a failing of the Lomb-Scargle \citep{Lomb76,Scargle82} periodogram which starts by subtracting the mean of the data and then fits a plain sinusoid. The floating mean periodogram works better than the Lomb-Scargle periodogram for small numbers of points. We obtain the $\chi^2$ of the fit as a function of $f$ and then identify minima in this function. 

In Table \ref{tab:results} we give the orbital parameters for each sdB binary. listing T$_{0}$, the systemic velocity, $\gamma$, the radial velocity semi-amplitude, K, and the reduced $\chi^2$ achieved for the best alias. We also give the period of an $N$th alias, and the difference in $\chi^2$ between this alias and the best alias. In most cases we list the $N = 2$nd alias, and find a significant $\chi^2$ difference between these best and second-best aliases. However, for some systems (PG0958-073, PG1000+408, PG1230+052, PG1403+316, PG1648+536, KPD2215+5037 and PG2331+038) we find that the best alias is surrounded by many other aliases which are very close in period and with a similar $\chi^2$. In some sense these systems are not solved since these close aliases are as significant as the best alias, but they are sufficiently close and span a sufficiently small range that our criterion is satisfied. For the purposes of Table \ref{tab:results} when the nearest competing aliases are so close in period it makes more sense to compare with the next {\it group} of aliases, so for these systems we choose to give the $N$th alias for which the period differs by more than $5$ per cent from the best alias. In all cases this results in a significant difference in $\chi^2$ between this alias and the best alias.

\begin{figure*}
\centering
\includegraphics[angle=270,width=1.0\textwidth]{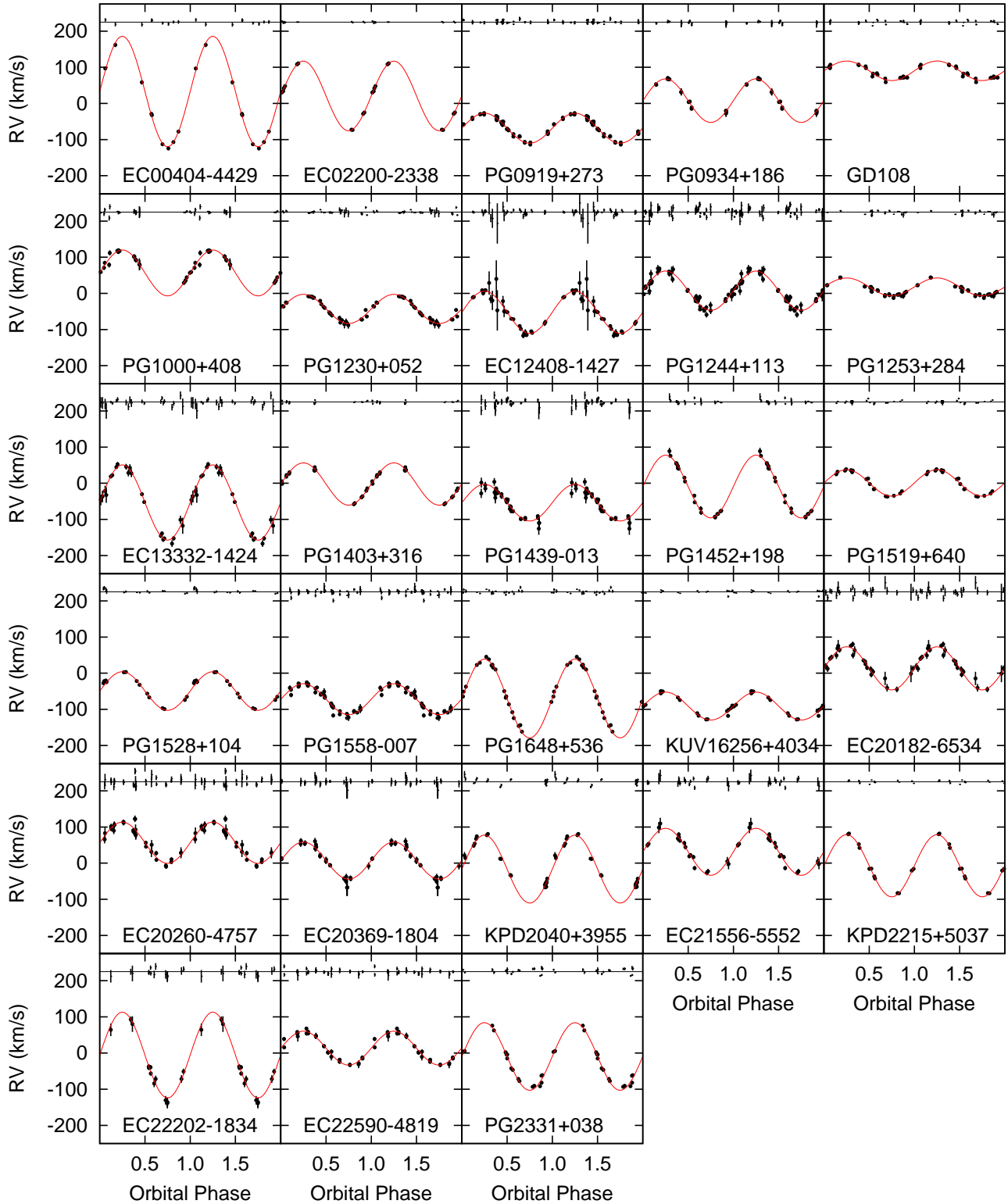}
\vspace{5mm}
\caption{The radial velocity curve for each object using the parameters for the best alias give in Table \ref{tab:results}, folded on the orbital period. For each object we also plot the residuals to the fit on the same scale, offset by $220$km/s.} \label{fig:phasefold} \end{figure*} 

The results of folding the radial velocities of each object on its orbital period are plotted in Figure \ref{fig:phasefold}. The error bars on the radial velocity points are, in most cases, smaller than the size of the symbol used to display them. 

\begin{figure*}
\centering
\includegraphics[angle=270,width=1.0\textwidth]{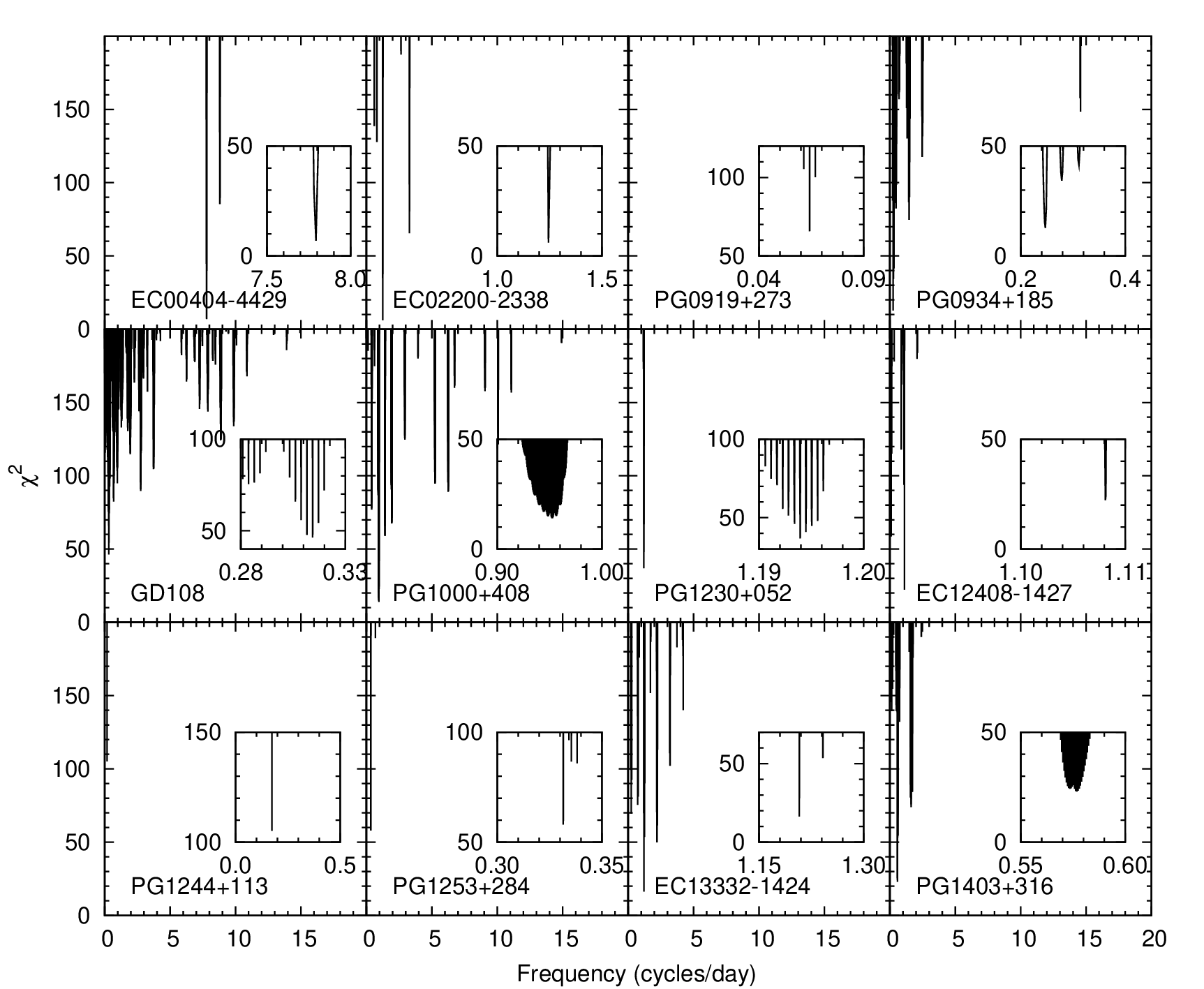}
\vspace{5mm}
\caption{Periodograms for the first 12 objects listed in Table \ref{tab:results}. Each panel presents $\chi^2$ versus cycles/day obtained after the period search was carried out. The frequency with the smallest $\chi^2$ corresponds to the orbital frequency of the system. For clarity we include an insert showing a blow-up of the region around the orbital frequency.} \label{fig:pgram1} \end{figure*}

\begin{figure*}
\centering
\includegraphics[angle=270,width=1.0\textwidth]{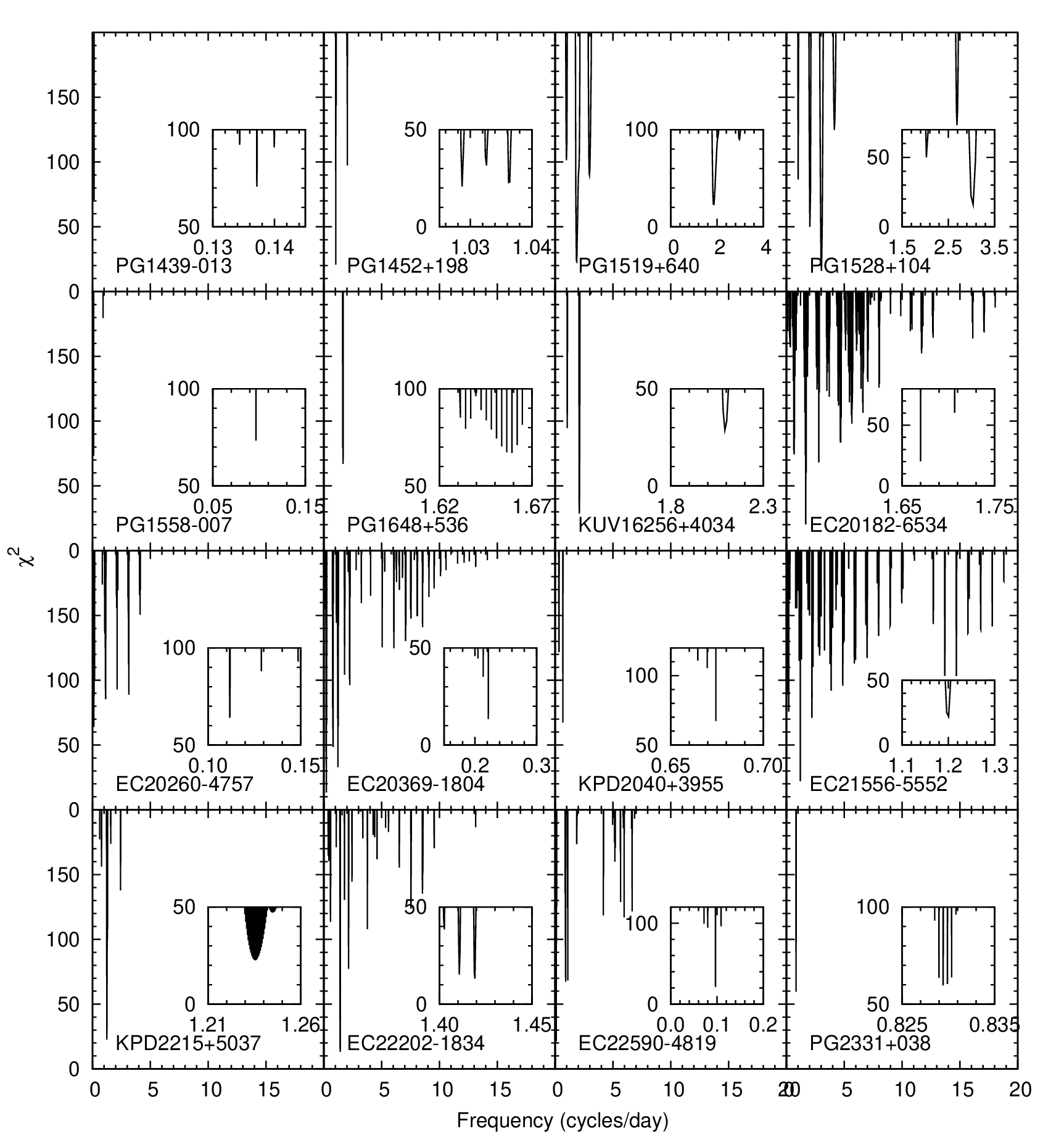}
\vspace{5mm}
\caption{Periodograms as in Figure \ref{fig:pgram1} for the remaining 16 objects listed in Table \ref{tab:results}.} \label{fig:pgram2} \end{figure*}

The periodograms ($\chi^2$ versus orbital frequency) for the 28 objects listed in Table \ref{tab:results} are given in Figures \ref{fig:pgram1} and \ref{fig:pgram2}. Each panel includes a
blow-up of the region in frequency where the minimum $\chi^2$ is found. It is clear from these figures that in the majority of cases there is a significant difference in $\chi^2$ between the best and the second alias. Exceptions are the seven systems we have previously discussed in which there are many aliases close in frequency and significance to the best alias. The blow-ups illustrate that the frequency range covered by these alternate aliases is very small, so we can determine the period to within $5$ per cent of the true value with confidence. Two other systems we wish to highlight are PG1452+198 and EC22202-1834. In Table \ref{tab:results} we compared the first and second aliases for these systems, which are very similar in significance. The periodograms for each of these two systems show that the two aliases are discrete and separate without the continuous range of intermediate aliases which we see in the previously discussed seven systems. For each system, either one of these two solutions could represent the `true' period, and the $\chi^2$ difference is too small to favour one over the other. However, in both cases the period difference between the two aliases is very small and our criterion for solution is satisfied no matter which we believe to be the true period. We therefore include these systems with those we consider to be solved.

\begin{table}
\caption{List of probabilities that the true orbital period of a
  system lies further than 1 and 5 per cent from our favoured
  value given in Table \ref{tab:results}. Numbers quoted are
  the logs in base 10 of the probabilities. Column number 4 gives the
  value of the systematic uncertainty that has been added in
  quadrature to the raw error to give a $\chi^2$ that lies above the
  2.5 per cent probability in the $\chi^2$ distribution.}
\label{tab:probs}
\begin{center}
\begin{tabular}{lrrc}
Object & 1\%\ & 5\%\  & systematic\\
& & &  error (km\,s$^{-1}$) \\
\hline
EC00404-4429	&$-15.09$   &$-15.09$   &2\\
EC02200-2338	&$-11.35$	&$-11.35$   &2\\
PG0919+273      &$-9.26$    &$-37.27$   &2\\
PG0934+186		&$-3.14$	&$-4.62$    &2\\
PG0958-073           &$-1.42$    &$-3.35$    &3\\
PG1000+408		&$-1.37$	&$-9.57$    &2\\
PG1230+052		&$-17.23$	&$-49.72$   &2\\
EC12408-1427	&$-7.54$	&$-7.54$    &2\\
PG1244+113		&$-25.47$	&$-27.97$   &4\\
PG1253+284      &$-2.70$    &$-14.94$   &3\\
EC13332-1424    &$-5.69$    &$-7.33$    &2\\
PG1403+316      &$-5.06$    &$-5.94$    &2\\
PG1439-013      &$-2.32$    &$-16.55$   &4\\
PG1452+198      &$-14.94$   &$-15.11$   &2\\
PG1519+640		&$-0.98$	&$-3.85$    &2\\
PG1528+104		&$-1.98$	&$-6.83$    &2\\
PG1558-007      &$-18.47$   &$-18.47$   &3\\
PG1648+536      &$-7.10$    &$-35.30$   &2\\
KUV16256+4034   &$-5.74$    &$-13.96$   &2\\
EC20182-6534    &$-8.57$    &$-10.04$   &2\\
EC20260-4757    &$-3.83$    &$-3.83$    &5\\
EC20369-1804    &$-4.40$    &$-4.83$    &2\\
KPD2040+3955	&$-2.75$	&$-7.31$    &4\\
EC21556-5552    &$-8.43$    &$-8.43$    &2\\
KPD2215+5037    &$-6.90$    &$-20.88$   &2\\
EC22202-1834    &$-6.48$    &$-9.21$    &2\\
EC22590-4819    &$-8.82$    &$-8.82$    &2\\
PG2331+038      &$-4.75$    &$-26.72$   &4\\ 
\end{tabular}
\end{center}
\end{table}

In Table~\ref{tab:probs} we list for each system the probability that the true orbital period lies further than 1 and 5 per cent from our favoured value, using the Bayesian calculation detailed in \citet{MoralesRueda03} and \citet{Marsh95}. We consider a period to be robust when the probability is below 0.1 per cent (or -3 in the log scale). This is not fulfilled to within 1 per cent of our favoured period for 7 of our 28 sources. The worst example is PG1519+640: for this system the probability that the true period is more than 1 per cent different from our favoured value is $-0.98$ in the log scale, or greater than 10 per cent. However, all of the periods are robust to within 5 per cent of our favoured values, and as we noted earlier this is the criterion by which we consider a system to be solved, since for the practical purpose of comparing the population of sdB stars to theoretical models knowing the periods to within 5 per cent is normally sufficient. In a number of cases the probability of the orbital period being further than 1 and 5 per cent from our favoured value is the same. This is because all the significant probability lies within a very small range around the best period, with all the significant competition (i.e. next best alias) placed outside the 5 per cent region around the best alias. 

We also compute the uncertainty that when added in quadrature to our raw error estimates gives a reduced $\chi^2 = 1$. This systematic uncertainty accounts for sources of errors such as true variability of the star or slit-filling errors. These errors are most probably not correlated with the orbit or the statistical errors determined and thus are added in quadrature. In all cases we use a minimum value of $2\,{\rm km}\,{\rm s}^{-1}$ corresponding to 1/10$^{\rm th}$ of a pixel which we believe to be a fair estimate of the true limits of our data. These determinations are also given in Table~\ref{tab:probs}. In most cases, the systematic uncertainty does not exceed the minimum value.

\section{Discussion}

\subsection{Effective temperature, surface gravity and helium abundance}
\label{sec:tefflogg}

\begin{table}
\caption{T$_{\rm eff}$, $\log g$ and $\log (\rm He/H)$. We list only values for the binary systems listed in Table \ref{tab:results} for which our data is sufficient to obtain a good constraint on these parameters.}
\label{tab:tefflogg}
\begin{center}
\begin{tabular}{llll}
Name & T$_{\rm eff}$ (K) & $\log g$ & $\log (\rm He/H)$ \\
\hline
PG0919+273      &$32900$      &$5.90$       &$-2.31$      \\
PG0934+186      &$35800$      &$5.65$       &$-3.00$      \\
PG1230+052      &$27100$      &$5.47$       &$-2.86$      \\
PG1244+113      &$36300$      &$5.54$       &$-3.00$      \\
PG1403+316      &$31200$      &$5.75$       &$-2.42$      \\
PG1452+198      &$29400$      &$5.75$       &$-2.00$      \\
PG1519+640      &$30600$      &$5.72$       &$-2.17$      \\
PG1528+104      &$27200$      &$5.46$       &$-2.52$      \\
PG1648+536      &$31400$      &$5.62$       &$-4.00$      \\
KUV16256+4034   &$23100$      &$5.38$       &$-2.99$      \\
KPD2040+3955    &$27900$      &$5.54$       &$-2.77$      \\
KPD2215+5037    &$29600$      &$5.64$       &$-2.24$      \\
PG2331+038      &$27200$      &$5.58$       &$-2.70$      \\\\
\end{tabular}
\end{center}
\end{table}

We measured the effective temperature, T$_{\rm eff}$, the surface gravity, $\log g$, and the helium abundance, $\log ({\rm He}/{\rm H})$, for 13 of the 28 sdBs listed in Table \ref{tab:results}, and list these measurements in Table~\ref{tab:tefflogg}. Due to the various instrument setups we used we are not able to do this for all the systems, because data in which the spectral range only encompasses a small number of lines is insufficient to constrain these parameters with any precision. We used the procedure of \citet{Saffer94} to fit the profiles of the Balmer, the He\,{\sc i} and the He\,{\sc ii} lines present in the spectra by a grid of synthetic spectra. The synthetic spectra obtained from the metal line-blanketed LTE model atmospheres of \citet{Heber00} were matched to the data simultaneously. For the two hottest stars the model grid with enhanced metal line blanketing was used, which substantially improved the fits (for details see \citealt{OToole06}). Before the fitting was carried out, we convolved the synthetic spectra with a Gaussian function to account for the instrumental profile. KPD2215+5037 was previously analysed with the same set of models in \citet{Heber02}, the results of which are in exact agreement with our finding here.

We plot these results in the T$_{\rm eff}$ / $\log g$ plane and find that all but two of the objects lie in the band defined by the zero-age extreme horizontal branch, the terminal-age extreme horizontal branch and the He main sequence (Figure 2 of \citealt{Maxted01}) and are therefore extreme horizontal branch (EHB) stars. The two exceptions are the two hottest stars, PG0934+186 and PG1244+113, which are sufficiently displaced from the band to be considered post-EHB stars. The effective temperature and gravity of PG1244+113 were determined by Saffer (as reported by \citealt{Maxted01}) with a method similar to ours but using a different grid of synthetic spectra to be T$_{\rm eff} = 33 800$K and $\log g = 5.67$, which places it just outside of the EHB band, although the uncertainty on the measurement was such that it could not be positively identified as a post-EHB star. We determine a a somewhat higher T$_{\rm eff}$ and lower gravity, which places the star away from the EHB. 

We also note that we find EC22133-6446 to be one of the rare helium-rich sdB stars. EC22133-6446 is one of the objects in our survey in which we detect no significant radial velocity variations.

\subsection{Orbital ellipticity}
\label{sec:eccentric}

\citet{Edelmann05} took high-resolution spectra of 15 sdB binaries, and for a third of these found the orbits to be slightly eccentric with $e \sim$$0.03$ -- $0.06$. To investigate ellipticity in our binaries we fitted each of our data sets using the Levenburg-Marquardt method \citep{Press02} with a model consisting of a sine function and its first harmonic, a reasonable approximation to an elliptical orbit for small values of $e$. The eccentricity is determined from the ratio of the amplitudes of the first harmonic to the fundamental. In Table \ref{tab:eccentric} we list the best-fit value of $e$ for each target, and the improvement in $\chi^{2}$ over a circular orbit. The change in the orbital period determination compared to the values reported in Table \ref{tab:results} is very small (less than one per cent in all cases, and much less than this in most). For each target, we computed the $F$ statistic comparing the elliptical model fit with the circular fit. In all cases, we find the improvement is not significant at the $95$\% level. We therefore find that there is no evidence of ellipticity in any of our radial velocity curves, not even at long periods where departures from circular orbits might be expected.

One important caveat is that our observations were not designed to detect ellipticity in these systems, and it is unlikely that our lower-resolution observations were capable of making significant detections of small eccentricities comparable to those reported by \citet{Edelmann05}. To investigate the limits of our data we used a Markov Chain Monte Carlo (MCMC) algorithm to perturb the elliptical orbit model parameters and compare to our radial velocity measurements. For each binary, we therefore determined the value of $e$ for which our data is sufficient to give a significant detection. We list the results in Table \ref{tab:eccentric}. These are the upper bounds on $e$, at the $68$, $95$ and $99$\% confidence levels. With the exception of PG0919+273, we find the upper bound on $e$ at the $95$ and $99$\% levels is much greater than the range reported by \citet{Edelmann05}. We therefore cannot rule out small eccentricities comparable to those reported by \citet{Edelmann05} in any of these binaries.

\begin{table}
\caption{For each binary, we list here the best-fit eccentricity ($e$) and the improvement in $\chi^{2}$ compared to the circular orbit fit. In all cases, the improvement is not significant at the $95$\% level. We also give the upper bounds on $e$ as determined from our MCMC calculation, at the $68$, $95$ and $99$\% confidence levels.}
\label{tab:eccentric}
\begin{center}
\begin{tabular}{llrlll}
Name & $e$ (best fit)&$ \Delta \chi^{2}$ &\multicolumn{3}{c}{upper bound on $e$} \\
     & &                   &$68$\% &$95$\% &$99$\%\\
\hline
EC00404-4429	&	0.06	&	5.6	&	0.08	&	0.11	&	0.12	\\
EC02200-2338	&	0.08	&	0.1	&	0.25	&	0.33	&	0.36	\\
PG0919+273	&	0.01	&	0.6	&	0.03	&	0.05	&	0.07	\\
PG0934+186	&	0.12	&	5.8	&	0.16	&	0.22	&	0.27	\\
PG0958-073	&	0.06	&	0.6	&	0.12	&	0.19	&	0.24	\\
PG1000+408	&	0.03	&	0.1	&	0.16	&	0.37	&	0.64	\\
PG1230+052	&	0.02	&	0.6	&	0.05	&	0.07	&	0.09	\\
EC12408-1427	&	0.06	&	1.3	&	0.08	&	0.14	&	0.17	\\
PG1244+113	&	0.01	&	0.2	&	0.05	&	0.08	&	0.10	\\
PG1253+284	&	0.13	&	7.5	&	0.16	&	0.23	&	0.27	\\
EC13332-1424	&	0.03	&	1.2	&	0.05	&	0.07	&	0.09	\\
PG1403+316	&	0.12	&	2.7	&	0.29	&	0.55	&	0.64	\\
PG1439-013	&	0.13	&	3.8	&	0.19	&	0.32	&	0.41	\\
PG1452+198	&	0.05	&	2.1	&	0.10	&	0.16	&	0.20	\\
PG1519+640	&	0.13	&	11.7	&	0.15	&	0.20	&	0.23	\\
PG1528+104	&	0.12	&	9.6	&	0.14	&	0.18	&	0.20	\\
PG1558-007	&	0.06	&	5.5	&	0.08	&	0.11	&	0.13	\\
PG1648+536	&	0.05	&	8.3	&	0.06	&	0.08	&	0.09	\\
KUV16256+4034	&	0.05	&	4.9	&	0.07	&	0.10	&	0.12	\\
EC20182-6534	&	0.06	&	1.5	&	0.10	&	0.16	&	0.19	\\
EC20260-4757	&	0.05	&	1.0	&	0.11	&	0.19	&	0.34	\\
EC20369-1804	&	0.03	&	0.2	&	0.09	&	0.16	&	0.21	\\
KPD2040+3955	&	0.13	&	11.7	&	0.15	&	0.22	&	0.27	\\
EC21556-5552	&	0.09	&	3.5	&	0.12	&	0.18	&	0.21	\\
KPD2215+5037	&	0.04	&	3.8	&	0.05	&	0.08	&	0.10	\\
EC22202-1834	&	0.15	&	4.5	&	0.19	&	0.27	&	0.33	\\
EC22590-4819	&	0.03	&	0.6	&	0.06	&	0.10	&	0.12	\\
PG2331+038	&	0.05	&	2.6	&	0.07	&	0.11	&	0.13	\\
\end{tabular}
\end{center}
\end{table}

\subsection{Unsolved systems, non-movers and the binary fraction}
\label{sec:binfrac}

In Table \ref{tab:rv2a} we list our radial velocity measurements for the 108 remaining sdBs which we have observed as a part of this project, but not previously published. In 88 of these systems, we detect no significant radial velocity variation. These systems are likely to be either single sdBs or binary systems in which the mass function is too low for us to detect radial velocity variations.

The remaining 20 sdBs do show significant radial velocity variations, and we consider it likely that most or all of these are binaries. Our current data are not sufficient to distinguish the true orbital period in these systems from various competing aliases, and so these targets are prime candidates for future measurements. We mark these systems in Table \ref{tab:rv2a} with an asterisk. Two systems of particular note are PG1610+519 and PG2317+046. The data we have collected to date suggest that the orbital periods of these systems may be relatively long at $50$ -- $70$ days.

This list of twenty candidate binaries is not exhaustive. As we have previously discussed, some fraction of the `single' sdBs may be long period systems. There are also a number of sdBs in Table \ref{tab:rv2a} for which we only obtained a small number of measurements and in which there may be large radial velocity variations which we have missed. There are also some systems in which a small number of measurements suggest a radial velocity variation which is formally significant, but when we fit them we find the only possible aliases are very low in amplitude and imply a mass function which is unphysical for an sdB with a stellar companion. We do not mark such systems as binary candidates. However, some authors have discussed the possibility of close substellar companions to sdB stars. One such star in our sample is HD149382. \citet{Geier09} inferred the presence of a close $8$--$23$ $M_J$ companion to this sdB using high-resolution radial velocity measurements, but the more recent data of \citet{Jacobs11} found no such variation. Our data for this star do show radial velocity variations which are formally significant. However, we do not believe these data support the presence of a close companion, for two reasons. Firstly, HD149382 has a $V$-band magnitude of $8.9$, making it by far the brightest star in our sample (the majority of our targets lie in the $12 > V > 14$ range). For such a bright star slit-filling errors are potentially larger than for faint stars because short exposures lead to larger deviations from the slit centre, and it is possible we have underestimated these in this case. Secondly, \citet{Jacobs11} observed a red companion star with a $1$'' separation ($\sim$$75$ AU) from the sdB. Given we used a $1$'' slit, it is possible that our measurements were contaminated by this distant companion. In general, we believe our intermediate dispersion measurements are not sufficient to make convincing claims for any substellar companions to the sdBs in our sample. 

As well as these 20 potential binaries and the 28 solved systems in this paper, we have previously published 23 binaries in \citet{Maxted02} and \citet{MoralesRueda03}. Additionally, there were 3 binaries which were originally on our target list but were since solved by other authors. Our best estimate for the binary fraction in the sdB population is therefore $46$ per cent. This is a lower bound for the reasons discussed in the previous paragraph. In addition, some binaries were already known in the Palomar-Green catalogue of sdBs before our project began. Since we may have included these systems in our sample had they not already been solved, our sample is slightly biased {\it against} binary systems. Including these systems in our binary fraction increases the estimate to $49$ per cent.

The initial study of the Palomar-Green sample of sdBs implied a binary fraction of $69\pm9$ per cent \citep{Maxted01}. However, the sdB binary fraction determined from the SPY (ESO Supernovae type Ia Progenitor surveY; \citealt{Napiwotzki01}) sdB sample was found to contain a binary fraction of $42$ per cent \citep{Napiwotzki04}. In comparing our binary fraction with the SPY result, we note that the SPY authors deliberately chose to exclude known composite spectrum objects from their survey. In Section \ref{sec:composite} we report 32 composite objects in our target list. When we exclude these objects, our binary fraction increases to $53$ -- $56$ per cent.

It was thought that the discrepancy between the \citep{Maxted01} and \citep{Napiwotzki04} binary fractions could be the result of the different populations surveyed in the two surveys, i.e. SPY looks at stars with white dwarf colours in the thick disk and halo whereas the Palomar-Green sample consists of targets with sdB colours in the thin disk. Our more extended study of the Palomar-Green sample (and the Edinburgh-Cape sample, which uses similar colour selection cuts) finds a binary fraction which is intermediate between the two earlier figures, and so we believe that the discrepancy is more likely due to low number statistics, rather than different populations being targeted. Finally, we note that the binary population synthesis of \citet{Han03} predicted an observable binary frequency of $55$ per cent for their best-fit simulation set (set 2) when selection effects are accounted for, which compares well with our finding.

\subsection{The nature of the companion stars}
\label{sec:composite}

\begin{table}
\caption{The minimum companion masses $M_{2 min}$ and mass functions $f_m$ for the 28 binary systems listed in Table \ref{tab:results}, both in units of \Msun. We also mark `WD' the seven systems for which the companion star has been determined to be a white dwarf by $^1$\citet{Maxted04} or $^2$\citet{Shimanskii08}.}
\label{tab:companionmass}
\begin{center}
\begin{tabular}{llll}
Object & $M_{2 min}$ & $f_m$ & Companion\\
\hline
EC00404-4429    &0.316  &0.047  &\\
EC02200-2338    &0.389  &0.074  &\\
PG0919+273      &0.480  &0.115  &\\
PG0934+186      &0.430  &0.092  &\\
PG0958-073      &0.142  &0.007  &\\
PG1000+408      &0.250  &0.028  &WD$^2$\\
PG1230+052      &0.132  &0.006  &WD$^1$\\
EC12408-1427    &0.212  &0.019  &\\
PG1244+113      &0.439  &0.096  &\\
PG1253+284      &0.123  &0.005  &\\
EC13332-1424    &0.441  &0.097  &\\
PG1403+316      &0.280  &0.036  &\\
PG1439-013      &0.445  &0.099  &\\
PG1452+198      &0.366  &0.065  &\\
PG1519+640      &0.100  &0.003  &WD$^1$\\
PG1528+104      &0.127  &0.005  &WD$^1$\\
PG1558-007      &0.412  &0.084  &\\
PG1648+536      &0.407  &0.082  &WD$^1$\\
KUV16256+4034   &0.101  &0.003  &WD$^1$\\
EC20182-6534    &0.183  &0.013  &\\
EC20260-4757    &0.589  &0.172  &\\
EC20369-1804    &0.362  &0.064  &\\
KPD2040+3955    &0.505  &0.128  &WD$^1$\\
EC21556-5552    &0.234  &0.024  &\\
KPD2215+5037    &0.333  &0.053  &\\
EC22202-1834    &0.494  &0.122  &\\
EC22590-4819    &0.470  &0.110  &\\
PG2331+038      &0.452  &0.102  &\\
\end{tabular}
\end{center}
\end{table}

We combine the orbital periods and the radial velocity semi-amplitudes listed in Table \ref{tab:results} to calculate the mass function, $f_m$, of the system according to the well-known relation:
\[f_m = \frac{M_2^3 \sin^3 i}{(M_1 +M_2)^2} = \frac{P K_{1}^3}{2 \pi G},\]
where the subscript `$1$' refers to the sdB star and `$2$' to its companion. If we take a canonical mass of $0.48$\Msun \ for the sdB star, we can also calculate the minimum mass of its companion, $M_{2\mathrm{min}}$. The values for $f_m$ and $M_{2\mathrm{min}}$ obtained in each case are given in Table \ref{tab:companionmass}. Additionally, we indicate seven systems for which the companion has been determined to be a white dwarf through the photometric studies of \citet{Maxted04} or \citet{Shimanskii08}, which showed an absence of reflection effects in these systems, implying a degenerate companion.

\begin{figure}
\centering
\includegraphics[angle=270,width=1.0\columnwidth]{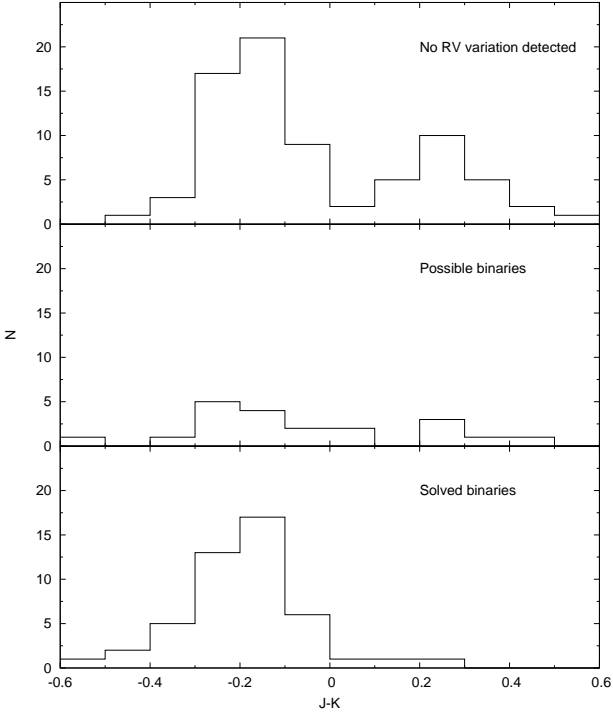}
\vspace{5mm}
\caption{Histogram of the J-K colours of the sdBs in our programme, obtained from 2MASS. We plot separately the 100 systems which show no significant radial velocity variation (top), the 25 systems which are candidate binaries but are currently unsolved (middle) and the solved binaries from this paper, \citet{Maxted02} and \citet{MoralesRueda03} (bottom).} \label{fig:histogram} \end{figure}

Some fraction of the sdB population are composite systems, in which flux excesses at long wavelengths or spectral features indicate the presence of a cool, G-K-type companion. The remaining `single' sdBs may truly be single stars, or they may have unseen white dwarf or fainter, dM-type companions. Using Two Micron All Sky Survey (2MASS)\footnote{At http://www.ipac.caltech.edu/2mass.} observations, \citet{Stark03} showed the composite and `single' sdB populations can be distinguished by their $J-K_s$ colour, with the `single' stars having $J-K_s < +0.05$ and the composites having $J-K_s > +0.05$. In figure 7 of \citet{Stark03} a histogram of the $J-K_s$ colours of all of the sdBs in the 2MASS Second Incremental Data Release showed a clear bimodal distribution. In Figure \ref{fig:histogram} we reproduce this histogram for the sdBs in our programme. We plot separately the sdBs which show no radial velocity variations, the sdBs which are strong binary candidates but without an orbital period determination, and the solved binaries, comprising the 28 systems from this paper and those previously published in \citet{Maxted02} and \citet{MoralesRueda03}. We exclude from this histogram the Kitt Peak-Downes (KPD; \citealt{Downes86}) survey objects, since they are close to the galactic plane and thus potentially significantly reddened.

If we examine first the histogram for the sdBs which shown no radial velocity variation, we see the same bimodal distribution around a $J-K_s$ value of $+0.05$ as was found in \citet{Stark03}. By comparison, the histogram of the solved binaries shows only two systems with $J-K_s > +0.05$: PG1253+284 and PG1558-007. PG1558-007 was identified as a composite system by \citet{Allard94}, but \citet{Heber02} disputed this identification, attributing the $J-K_s$ colour to interstellar reddening. PG1253+284 was identified as a composite system by \citet{Ulla98}, but for this system \citet{Heber02} determined PG1253+284 (referred to as TON 139 in that paper) to be a triple system via \hst \ imaging, and concluded that the $J-K_s$ colour in this system is due to the third, distant component, and not the companion in the close binary. The third histogram (the potential but unsolved binaries) indicates the presence of a further five composite systems. Aside from the possibility that these are close binaries with a G-K companion, there are three explanations for these measurements. Firstly, some of these composite sdBs may not actually be close binaries: further observations may show the radial velocity variations detected to date are not significant. Secondly, some of the $J-K_s$ measurements may be due to a third component, interstellar reddening or a nearby unresolved field star. Thirdly, some of these unsolved binaries may actually be long period systems, as we noted in Section \ref{sec:binfrac}.

To investigate this further, we generated mean spectra for all of the objects in our target list, and classified them with the aid of synthetic spectra from the grid described in Section \ref{sec:tefflogg}. 25 objects show contamination in their spectra indicative of a cool companion, which would indicate these are composite systems. We would classify these sdBs as `double-lined spectroscopic binaries', to distinguish them from the single-lined spectroscopic binaries, the nature of which we identify via radial velocity variations. There is strong overlap between the double-lined spectroscopic binaries and the composite systems we identified through 2MASS, with all but five of the 2MASS systems showing a contaminating component in their spectra. We list our candidate composite binaries in Table \ref{tab:composite}. We exclude PG2059+013 from this table: while the $J-K_s$ colour of this sdB is consistent with it being a composite system, but the \citet{Schlegel98} dust maps show it to be significantly reddened. This table contains five systems in which we believe we detect significant radial velocity variations, and hence are potentially close binaries.

In summary then, almost all of the systems which we identify as close binaries do not show the presence of a dwarf G-K companion. The companions in these systems are therefore most likely white dwarfs or M dwarfs. The composite systems are almost entirely found in the group of sdBs in which we detected no radial velocity variations. A G-K companion therefore seems to be indicative of a longer period system: a wide binary in which the sdB has evolved independently of the companion. There are some potential exceptions to this rule: these sdBs are prime candidates for further observation in order to determine if they are indeed binaries, as the data collected to date would suggest, and furthermore if they are close binaries. As we remarked in Section \ref{sec:binfrac}, two of these candidates (PG1610+519 and PG2317+046) show evidence for binarity, but all of the current aliases suggest a long orbital period, which would be consistent with our composite identification.

\begin{table}

\caption{sdBs in our sample which show evidence for being composite systems. All of the sdBs we list have 2MASS colours which indicate a cool companion. The majority of these systems are `double-lined' binaries, in which the inspection of our own spectra show evidence for a contaminating component. The four systems we list at the end show no such evidence. We also embolden the five objects which show some evidence for binarity through radial velocity variations.}
\label{tab:composite}
\begin{center}
\begin{tabular}{ll}
PG0039+049              &EC20117-4014\\
PB8783                  &{\bf EC20228-3628}\\
EC03143-5945            &EC20387-1716\\
EC03238-0710            &EC21079-3548\\
EC04170-3433            &PG2110+127\\
{\bf EC04515-3754}      &PG2118+126\\
EC05053-2806            &EC21373-3727\\
PG0749+658              &PG2148+095\\
EC09436-0929            &PG2226+094\\
PG1040+234              &\\
EC12473-3046            &\bf\underline{2MASS only}\\
PG1338+611              &EC03470-5039 \\
PG1551+256              &{\bf PG1526+132} \\
{\bf PG1610+519}        &EC22133-6446  \\
PG1618+563              &{\bf PG2317+046} \\
PG1701+359              &\\
\end{tabular}
\end{center}
\end{table}

\subsection{Misclassifications in the literature}

Our main source for the construction of our target list was \citet{Kilkenny88}, with the Edinburgh Cape objects coming from \citet{Kilkenny97} and private communications. Over the course of our study we discovered a number of sources which have been misidentified as sdB stars in these catalogues, which we list in Table \ref{tab:misclassifications}. At the time of writing they are all still listed as sdBs in the SIMBAD astronomical database, although the misidentification has previously been reported for six of these eleven objects. For the other five we give our new classifications, obtained using the same spectral fitting techniques described in Section \ref{sec:tefflogg}.

\begin{table}
\caption{Ten objects in our survey which were misidentified as sdB stars in \citet{Kilkenny88} or \citet{Kilkenny97}. Six of these objects have previously been reclassified by $^1$\citet{Gianninas10}, $^2$\citet{Saffer97} or $^3$\citet{Ramspeck01}.}
\label{tab:misclassifications}
\begin{center}
\begin{tabular}{ll}
Object & Our classification\\
\hline
KPD~0311+4801   & DA, T$_{\rm eff} = 97 080$K, $\log g = 6.96$ $^1$\\
KUV~04110+1434  & MS B-star,  T$_{\rm eff} = 13 000$K\\
UVO0653-23      & B-star\\
KPD~0721-0026   & B-star, T$_{\rm eff} = 11 868$K, $\log g = 3.68$ $^2$\\
HD~80836        & MS B-star\\
EC10282-1605    & MS B-star,  T$_{\rm eff} \sim 16 000$K\\
EC13506-3137    & sdO star of unusually low gravity\\
PG1533+467      & MS B-star, T$_{\rm eff} = 18 100$K, $\log g = 4.00$ $^3$\\
KPD~2022+2033   & B-star, T$_{\rm eff} = 16 752$K, $\log g = 4.46$ $^2$\\
PG2111+023      & MS B-star, T$_{\rm eff} = 18 305$K, $\log g = 4.5$ $^2$\\
PG2301+259      & MS B-star, T$_{\rm eff} = 17 901$K, $\log g = 4.11$ $^2$\\
\end{tabular}
\end{center}
\end{table}

\section{CONCLUSIONS}

In this paper we present a large number of radial velocity measurements of bright sdB stars. The aim of this project was to detect binary systems through variations in these radial velocity measurements, and hence determine the orbital parameters of those systems. This is a continuation of the work begun in \citet{Maxted02} and \citet{MoralesRueda03}. We presented a total of 28 new binary systems with their parameters, in all cases determining the orbital period to much better than $5$ per cent. We determined effective temperatures and surface gravities for 13 of the sdBs in these systems. The parameters are consistent with the sdBs being extreme horizontal branch stars (EHB) with two exceptions, PG0934+186 and PG1244+113, which we classify as post-EHB stars.

As well as the 28 solved systems we presented measurements for 108 other sdBs. 20 of these show strong signs of binarity, but our data is insufficient as yet to constrain the system parameters. The remaining 100 stars show no significant radial velocity variations and are likely to be either single sdBs or binary systems in which the orbital period is long enough to push the radial velocity variations below our detection threshold. Our best estimate for the binary fraction is $46$ -- $56$ per cent. Finally, we note that none of the binaries we report in this paper show the near-infrared colours which indicate the presence of a G-K-type companion star. The companion stars in our solved systems are therefore likely either white dwarfs or M dwarfs. There is some evidence for a cool companion in the spectra of 32 of the remaining sdBs, including 5 which show radial velocity variations. These are strong candidates for future study.

\section*{ACKNOWLEDGEMENTS}
CMC and TRM are supported under grant ST/F002599/1 from the Science and Technology Facilities Council (STFC). The results presented in this paper are based on observations made with the Isaac Newton Telescope operated on the island of La Palma by the Isaac Newton Group in the Spanish Observatorio del Roque de los Muchachos of the Institutio de Astrofisica de Canarias  and observations made with the 1.9 m telescope operated by the South African Astronomical Observatory. This research has made use of NASA's Astrophysics Data System Bibliographic Services and the SIMBAD database, operated at CDS, Strasbourg, France. This publication makes use of data products from the Two Micron All Sky Survey, which is a joint project of the University of Massachusetts and the Infrared Processing and Analysis Center/California Institute of Technology, funded by the National Aeronautics and Space Administration and the National Science Foundation. We thank the referee for helpful comments.

\bibliography{sdb}

\begin{table*}
  \caption{Radial velocity measurements for the 28 sdB binaries which we present in this paper.}
  \label{tab:rv1a}
  \begin{center}
\begin{tabular}{lr@{\,$\pm$\,}llr@{\,$\pm$\,}llr@{\,$\pm$\,}llr@{\,$\pm$\,}l}
HMJD & \multicolumn{2}{c}{RV}         & HMJD & \multicolumn{2}{c}{RV}         & HMJD & \multicolumn{2}{c}{RV}         & HMJD & \multicolumn{2}{c}{RV}\\
     & \multicolumn{2}{c}{(km s$^{-1}$)} &      & \multicolumn{2}{c}{(km s$^{-1}$)} &      & \multicolumn{2}{c}{(km s$^{-1}$)} &      & \multicolumn{2}{c}{(km s$^{-1}$)}\\
\hline    
\multicolumn{3}{c}{\bf\underline{EC00404-4429}}					&	\multicolumn{3}{c}{\bf\underline{PG0919+273}}					&	\multicolumn{3}{c}{\bf\underline{PG1000+408}}					&	\multicolumn{3}{c}{\bf\underline{PG1244+113}}					\\
52892.9858	&	-124.51	&	5.13	&	54902.9529	&	-74.14	&	1.26	&	54191.9382	&	28.74	&	2.11	&	51654.1003	&	-21.09	&	5.70	\\
52893.0000	&	-78.05	&	5.11	&	54902.9704	&	-74.38	&	1.26	&	54191.9522	&	33.23	&	2.00	&	51977.1722	&	-32.37	&	9.29	\\
52893.1529	&	96.73	&	4.86	&	\multicolumn{3}{c}{\bf\underline{PG0934+186}}					&	\multicolumn{3}{c}{\bf\underline{PG1230+052}}					&	51977.1827	&	-48.06	&	9.30	\\
52893.1671	&	161.87	&	5.97	&	52360.9826	&	67.29	&	3.19	&	52387.9376	&	-85.66	&	11.24	&	51982.1343	&	-20.65	&	6.05	\\
52893.9884	&	-28.82	&	4.12	&	52360.9966	&	70.62	&	3.21	&	52387.9534	&	-78.37	&	8.46	&	51982.1448	&	-17.57	&	5.95	\\
\end{tabular}
  \end{center}
\end{table*}
\setcounter{table}{8}
\begin{table*}
  \caption{Radial velocity measurements for the 28 sdB binaries which we present in this paper.}
  \label{tab:rv1b}
  \begin{center}
    \begin{tabular}{lr@{\,$\pm$\,}llr@{\,$\pm$\,}llr@{\,$\pm$\,}llr@{\,$\pm$\,}l}
HMJD & \multicolumn{2}{c}{RV}         & HMJD & \multicolumn{2}{c}{RV}         & HMJD & \multicolumn{2}{c}{RV}         & HMJD & \multicolumn{2}{c}{RV}\\
     & \multicolumn{2}{c}{(km s$^{-1}$)} &      & \multicolumn{2}{c}{(km s$^{-1}$)} &      & \multicolumn{2}{c}{(km s$^{-1}$)} &      & \multicolumn{2}{c}{(km s$^{-1}$)}\\
\hline    
\multicolumn{3}{c}{\bf\underline{PG1253+284}}					&	\multicolumn{3}{c}{\bf\underline{PG1439-013}}					&	\multicolumn{3}{c}{\bf\underline{PG1519+640}}					&	\multicolumn{3}{c}{\bf\underline{PG1558-007}}					\\
54190.1183	&	43.52	&	1.08	&	52360.0982	&	-109.88	&	15.03	&	52390.9823	&	38.86	&	2.47	&	53103.0441	&	-99.63	&	2.87	\\
54191.1132	&	15.61	&	0.85	&	52363.0472	&	-5.19	&	9.27	&	52391.0106	&	35.96	&	4.38	&	53103.0651	&	-101.28	&	2.96	\\
54191.1255	&	17.70	&	1.00	&	52363.0578	&	-14.69	&	10.56	&	52391.0143	&	35.17	&	5.33	&	53543.9387	&	-60.22	&	8.61	\\
54193.9361	&	18.93	&	0.91	&	52719.9969	&	-28.05	&	12.92	&	52391.0180	&	30.89	&	5.72	&	53543.9531	&	-51.02	&	9.63	\\
54193.9501	&	19.01	&	0.91	&	52720.0110	&	1.88	&	13.70	&	52391.0217	&	34.95	&	4.61	&	53544.8236	&	-88.18	&	6.64	\\
\end{tabular}
  \end{center}
\end{table*}
\setcounter{table}{8}
\begin{table*}
  \caption{Radial velocity measurements for the 28 sdB binaries which we present in this paper.}
  \label{tab:rv1c}
  \begin{center}
    \begin{tabular}{lr@{\,$\pm$\,}llr@{\,$\pm$\,}llr@{\,$\pm$\,}llr@{\,$\pm$\,}l}
HMJD & \multicolumn{2}{c}{RV}         & HMJD & \multicolumn{2}{c}{RV}         & HMJD & \multicolumn{2}{c}{RV}         & HMJD & \multicolumn{2}{c}{RV}\\
     & \multicolumn{2}{c}{(km s$^{-1}$)} &      & \multicolumn{2}{c}{(km s$^{-1}$)} &      & \multicolumn{2}{c}{(km s$^{-1}$)} &      & \multicolumn{2}{c}{(km s$^{-1}$)}\\
\hline 
\multicolumn{3}{c}{\bf\underline{EC20182-6534}}					&	\multicolumn{3}{c}{\bf\underline{EC20369-1804}}					&	\multicolumn{3}{c}{\bf\underline{EC21556-5552}}					&	\multicolumn{3}{c}{\bf\underline{EC22590-4819}}					\\
53303.8700	&	68.48	&	7.28	&	52893.9340	&	-6.77	&	3.96	&	53655.9307	&	79.46	&	6.34	&	52896.8993	&	54.81	&	7.33	\\
53303.8841	&	64.39	&	7.20	&	52895.8477	&	-7.98	&	9.60	&	53655.9414	&	62.31	&	6.25	&	52896.9100	&	45.90	&	14.20	\\
53543.9901	&	49.05	&	9.75	&	53300.8000	&	-40.33	&	5.98	&	53656.8616	&	19.04	&	5.09	&	52897.9100	&	67.73	&	4.92	\\
53544.0012	&	75.57	&	16.40	&	53300.8108	&	-43.49	&	6.39	&	53656.8723	&	22.31	&	4.95	&	52897.9207	&	54.27	&	5.96	\\
53546.0133	&	11.16	&	8.18	&	53301.8784	&	11.87	&	5.51	&	53657.8573	&	-26.69	&	6.93	&	53663.7933	&	53.59	&	6.61	\\
\end{tabular}
  \end{center}
\end{table*}


\setcounter{table}{9}
\begin{table*}
  \caption{Previously unpublished radial velocities for the remaining sdBs in our survey. Some of these show no significant variation. Others (which we mark with an asterisk) show signs of orbital variations but our data is insufficient to distinguish between competing aliases.}
  \label{tab:rv2a}
  \begin{center}
    \begin{tabular}{lr@{\,$\pm$\,}llr@{\,$\pm$\,}llr@{\,$\pm$\,}llr@{\,$\pm$\,}l}
HMJD & \multicolumn{2}{c}{RV}         & HMJD & \multicolumn{2}{c}{RV}         & HMJD & \multicolumn{2}{c}{RV}         & HMJD & \multicolumn{2}{c}{RV}\\
     & \multicolumn{2}{c}{(km s$^{-1}$)} &      & \multicolumn{2}{c}{(km s$^{-1}$)} &      & \multicolumn{2}{c}{(km s$^{-1}$)} &      & \multicolumn{2}{c}{(km s$^{-1}$)}\\
\hline    
\multicolumn{3}{c}{\bf\underline{EC00042-2737}}					&	\multicolumn{3}{c}{\bf\underline{EC00292-4741*}}					&	\multicolumn{3}{c}{\bf\underline{KPD0054+5406}}					&	\multicolumn{3}{c}{\bf\underline{EC03143-5945}}					\\
53303.9894	&	33.51	&	9.34	&	53658.0042	&	-8.25	&	5.41	&	52132.1353	&	-37.95	&	3.30	&	53659.0314	&	50.50	&	8.85	\\
53304.0040	&	30.00	&	10.31	&	53658.8577	&	24.01	&	7.21	&	52183.1818	&	-40.66	&	2.45	&	\multicolumn{3}{c}{\bf\underline{EC03238-0710}}					\\
53547.1765	&	39.25	&	8.63	&	53658.8649	&	12.75	&	8.38	&	52183.1958	&	-39.09	&	2.46	&	53300.0649	&	-10.38	&	5.53	\\
53657.9673	&	36.81	&	11.90	&	53663.9137	&	16.72	&	2.87	&	52184.1074	&	-27.10	&	4.12	&	53300.0721	&	12.15	&	5.62	\\
53657.9815	&	45.64	&	11.38	&	53665.7949	&	19.23	&	4.01	&	52184.1215	&	-38.24	&	3.12	&	53301.0418	&	-0.11	&	4.88	\\
\end{tabular}
  \end{center}
\end{table*}

\setcounter{table}{9}
\begin{table*}
  \caption{Previously unpublished radial velocities for the remaining sdBs in our survey. Some of these show no significant variation. Others (which we mark with an asterisk) show signs of orbital variations but our data is insufficient to distinguish between competing aliases.}
  \label{tab:rv2b}
  \begin{center}
    \begin{tabular}{lr@{\,$\pm$\,}llr@{\,$\pm$\,}llr@{\,$\pm$\,}llr@{\,$\pm$\,}l}
HMJD & \multicolumn{2}{c}{RV}         & HMJD & \multicolumn{2}{c}{RV}         & HMJD & \multicolumn{2}{c}{RV}         & HMJD & \multicolumn{2}{c}{RV}\\
     & \multicolumn{2}{c}{(km s$^{-1}$)} &      & \multicolumn{2}{c}{(km s$^{-1}$)} &      & \multicolumn{2}{c}{(km s$^{-1}$)} &      & \multicolumn{2}{c}{(km s$^{-1}$)}\\
\hline    
\multicolumn{3}{c}{\bf\underline{EC04170-3433}}					&	\multicolumn{3}{c}{\bf\underline{PG0749+658}}					&	\multicolumn{3}{c}{\bf\underline{PG0749+658}}					&	\multicolumn{3}{c}{\bf\underline{PG0749+658}}					\\
53658.1051	&	69.00	&	1.90	&	51945.8625	&	-13.11	&	1.31	&	52036.8594	&	-6.57	&	2.84	&	54185.8518	&	-10.13	&	2.91	\\
53659.0714	&	68.94	&	2.84	&	51976.8323	&	-8.10	&	1.18	&	52036.8601	&	-8.38	&	2.86	&	54190.8678	&	-18.49	&	2.43	\\
53659.0752	&	69.42	&	2.90	&	51976.8358	&	-9.70	&	1.13	&	52036.8608	&	-16.12	&	2.85	&	54190.8701	&	-16.25	&	2.51	\\
53662.0616	&	71.37	&	2.76	&	51977.8255	&	-15.65	&	1.62	&	52036.8616	&	-5.13	&	2.78	&	\multicolumn{3}{c}{\bf\underline{PG0909+164}}					\\
53662.0653	&	75.51	&	3.12	&	51977.8291	&	-15.63	&	1.48	&	52036.8623	&	-11.33	&	2.86	&	52361.8937	&	51.79	&	3.54	\\
\end{tabular}
  \end{center}
\end{table*}

\setcounter{table}{9}
\begin{table*}
  \caption{Previously unpublished radial velocities for the remaining sdBs in our survey. Some of these show no significant variation. Others (which we mark with an asterisk) show signs of orbital variations but our data is insufficient to distinguish between competing aliases.}
  \label{tab:rv2c}
  \begin{center}
    \begin{tabular}{lr@{\,$\pm$\,}llr@{\,$\pm$\,}llr@{\,$\pm$\,}llr@{\,$\pm$\,}l}
HMJD & \multicolumn{2}{c}{RV}         & HMJD & \multicolumn{2}{c}{RV}         & HMJD & \multicolumn{2}{c}{RV}         & HMJD & \multicolumn{2}{c}{RV}\\
     & \multicolumn{2}{c}{(km s$^{-1}$)} &      & \multicolumn{2}{c}{(km s$^{-1}$)} &      & \multicolumn{2}{c}{(km s$^{-1}$)} &      & \multicolumn{2}{c}{(km s$^{-1}$)}\\
\hline    
\multicolumn{3}{c}{\bf\underline{EC10189-1502}}					&	\multicolumn{3}{c}{\bf\underline{PG1040+234}}					&	\multicolumn{3}{c}{\bf\underline{EC12376-2049}}					&	\multicolumn{3}{c}{\bf\underline{PG1338+611}}					\\
53101.8854	&	-27.59	&	12.37	&	54906.0269	&	15.01	&	1.53	&	53545.7824	&	30.32	&	11.96	&	54195.2702	&	44.56	&	1.94	\\
53101.8961	&	-19.20	&	12.14	&	54906.9538	&	13.33	&	1.42	&	53548.7453	&	43.27	&	11.97	&	54195.2745	&	-10.14	&	3.71	\\
53101.9085	&	4.20	&	12.26	&	54906.9713	&	10.77	&	1.40	&	53548.7597	&	57.06	&	12.35	&	54195.8929	&	42.56	&	0.88	\\
53101.9192	&	16.63	&	12.05	&	\multicolumn{3}{c}{\bf\underline{PG1047+003}}					&	\multicolumn{3}{c}{\bf\underline{EC12473-3046}}					&	54195.9035	&	43.86	&	0.87	\\
53101.9298	&	8.18	&	12.43	&	51976.9999	&	-11.38	&	3.48	&	53545.8062	&	76.64	&	6.42	&	54196.9553	&	45.27	&	1.26	\\
\end{tabular}
  \end{center}
\end{table*}

\setcounter{table}{9}
\begin{table*}
  \caption{Previously unpublished radial velocities for the remaining sdBs in our survey. Some of these show no significant variation. Others (which we mark with an asterisk) show signs of orbital variations but our data is insufficient to distinguish between competing aliases.}
  \label{tab:rv2d}
  \begin{center}
    \begin{tabular}{lr@{\,$\pm$\,}llr@{\,$\pm$\,}llr@{\,$\pm$\,}llr@{\,$\pm$\,}l}
HMJD & \multicolumn{2}{c}{RV}         & HMJD & \multicolumn{2}{c}{RV}         & HMJD & \multicolumn{2}{c}{RV}         & HMJD & \multicolumn{2}{c}{RV}\\
     & \multicolumn{2}{c}{(km s$^{-1}$)} &      & \multicolumn{2}{c}{(km s$^{-1}$)} &      & \multicolumn{2}{c}{(km s$^{-1}$)} &      & \multicolumn{2}{c}{(km s$^{-1}$)}\\
\hline    
\multicolumn{3}{c}{\bf\underline{PG1417+257}}					&	\multicolumn{3}{c}{\bf\underline{PG1432+004}}					&	\multicolumn{3}{c}{\bf\underline{PG1526+132*}}					&	\multicolumn{3}{c}{\bf\underline{PG1610+519*}}					\\
52393.0569	&	-5.14	&	2.05	&	52361.1965	&	-0.79	&	1.62	&	52362.1744	&	5.54	&	4.88	&	52187.8507	&	-19.15	&	2.39	\\
52393.0927	&	-7.77	&	2.50	&	52362.1548	&	2.35	&	1.99	&	52362.1884	&	11.92	&	4.71	&	52748.0400	&	13.59	&	5.00	\\
52393.1033	&	-9.47	&	2.55	&	52362.1619	&	0.88	&	1.91	&	52362.2296	&	3.83	&	2.98	&	52748.0541	&	10.02	&	4.30	\\
52739.0635	&	0.05	&	3.16	&	52363.2310	&	2.15	&	3.69	&	52362.2436	&	2.98	&	2.64	&	52748.1636	&	3.96	&	4.82	\\
52739.0742	&	-3.81	&	3.17	&	52363.2418	&	1.52	&	3.60	&	54190.1838	&	10.59	&	3.03	&	52748.1777	&	6.03	&	5.68	\\
\end{tabular}
  \end{center}
\end{table*}

\setcounter{table}{9}
\begin{table*}
  \caption{Previously unpublished radial velocities for the remaining sdBs in our survey. Some of these show no significant variation. Others (which we mark with an asterisk) show signs of orbital variations but our data is insufficient to distinguish between competing aliases.}
  \label{tab:rv2e}
  \begin{center}
    \begin{tabular}{lr@{\,$\pm$\,}llr@{\,$\pm$\,}llr@{\,$\pm$\,}llr@{\,$\pm$\,}l}
HMJD & \multicolumn{2}{c}{RV}         & HMJD & \multicolumn{2}{c}{RV}         & HMJD & \multicolumn{2}{c}{RV}         & HMJD & \multicolumn{2}{c}{RV}\\
     & \multicolumn{2}{c}{(km s$^{-1}$)} &      & \multicolumn{2}{c}{(km s$^{-1}$)} &      & \multicolumn{2}{c}{(km s$^{-1}$)} &      & \multicolumn{2}{c}{(km s$^{-1}$)}\\
\hline    
\multicolumn{3}{c}{\bf\underline{HD149382}}					&	\multicolumn{3}{c}{\bf\underline{HD149382}}					&	\multicolumn{3}{c}{\bf\underline{PG1701+359}}					&	\multicolumn{3}{c}{\bf\underline{KPD1924+2932}}					\\
52364.1566	&	17.98	&	1.63	&	52393.0285	&	32.03	&	1.27	&	54188.2578	&	-120.96	&	1.88	&	52183.9450	&	57.44	&	2.98	\\
52364.1591	&	17.84	&	1.76	&	52393.0301	&	32.59	&	1.27	&	54188.2722	&	-125.33	&	3.68	&	52183.9521	&	60.81	&	3.63	\\
52364.1616	&	17.55	&	1.74	&	52393.0318	&	29.37	&	1.23	&	54189.1522	&	-124.95	&	1.31	&	52186.0000	&	45.92	&	4.91	\\
52364.1642	&	17.58	&	1.64	&	52739.2503	&	24.72	&	0.93	&	54189.1662	&	-123.32	&	1.29	&	52186.0071	&	62.25	&	4.66	\\
52364.1667	&	19.34	&	1.60	&	52739.2523	&	22.69	&	1.09	&	54330.1793	&	-114.28	&	1.16	&	52186.9886	&	51.84	&	3.31	\\
\end{tabular}
  \end{center}
\end{table*}

\setcounter{table}{9}
\begin{table*}
  \caption{Previously unpublished radial velocities for the remaining sdBs in our survey. Some of these show no significant variation. Others (which we mark with an asterisk) show signs of orbital variations but our data is insufficient to distinguish between competing aliases.}
  \label{tab:rv2f}
  \begin{center}
    \begin{tabular}{lr@{\,$\pm$\,}llr@{\,$\pm$\,}llr@{\,$\pm$\,}llr@{\,$\pm$\,}l}
HMJD & \multicolumn{2}{c}{RV}         & HMJD & \multicolumn{2}{c}{RV}         & HMJD & \multicolumn{2}{c}{RV}         & HMJD & \multicolumn{2}{c}{RV}\\
     & \multicolumn{2}{c}{(km s$^{-1}$)} &      & \multicolumn{2}{c}{(km s$^{-1}$)} &      & \multicolumn{2}{c}{(km s$^{-1}$)} &      & \multicolumn{2}{c}{(km s$^{-1}$)}\\
\hline    
\multicolumn{3}{c}{\bf\underline{EC20117-4014}}					&	\multicolumn{3}{c}{\bf\underline{PG2059+013}}					&	\multicolumn{3}{c}{\bf\underline{KPD2118+3841}}					&	\multicolumn{3}{c}{\bf\underline{PG2223+143}}					\\
53657.8384	&	12.15	&	3.07	&	54335.1524	&	-12.64	&	3.75	&	52188.0969	&	-4.91	&	3.38	&	54342.1777	&	-2.24	&	0.49	\\
\multicolumn{3}{c}{\bf\underline{EC20228-3628*}}					&	54335.9628	&	-15.97	&	3.99	&	\multicolumn{3}{c}{\bf\underline{PG2118+126}}					&	54342.1917	&	0.59	&	0.49	\\
52892.8803	&	-72.71	&	5.28	&	54335.9803	&	-9.93	&	4.04	&	52131.9309	&	-4.44	&	2.25	&	54342.9286	&	-7.58	&	0.42	\\
52892.8947	&	-82.57	&	4.39	&	54337.0496	&	-8.52	&	4.41	&	52131.9415	&	-6.37	&	2.25	&	54342.9427	&	-8.95	&	0.47	\\
52893.8655	&	-71.12	&	2.72	&	54337.0671	&	-4.94	&	3.86	&	52188.0709	&	3.06	&	2.30	&	54343.0569	&	-5.47	&	0.45	\\
\end{tabular}
  \end{center}
\end{table*}

\setcounter{table}{9}
\begin{table*}
  \caption{Previously unpublished radial velocities for the remaining sdBs in our survey. Some of these show no significant variation. Others (which we mark with an asterisk) show signs of orbital variations but our data is insufficient to distinguish between competing aliases.}
  \label{tab:rv2h}
  \begin{center}
    \begin{tabular}{lr@{\,$\pm$\,}llr@{\,$\pm$\,}llr@{\,$\pm$\,}llr@{\,$\pm$\,}l}
HMJD & \multicolumn{2}{c}{RV}         & HMJD & \multicolumn{2}{c}{RV}         & HMJD & \multicolumn{2}{c}{RV}         & HMJD & \multicolumn{2}{c}{RV}\\
     & \multicolumn{2}{c}{(km s$^{-1}$)} &      & \multicolumn{2}{c}{(km s$^{-1}$)} &      & \multicolumn{2}{c}{(km s$^{-1}$)} &      & \multicolumn{2}{c}{(km s$^{-1}$)}\\
\hline    
\multicolumn{3}{c}{\bf\underline{PG2314+076*}}					&	\multicolumn{3}{c}{\bf\underline{PG2314+076*}}					&	\multicolumn{3}{c}{\bf\underline{PG2317+046*}}					&	\multicolumn{3}{c}{\bf\underline{EC23483-6445}}					\\
52133.0195	&	-31.33	&	2.54	&	54341.1389	&	-39.29	&	2.20	&	52182.1409	&	-42.77	&	2.00	&	53301.9581	&	13.18	&	4.65	\\
52133.0335	&	-31.14	&	2.50	&	54341.1599	&	-38.16	&	1.92	&	52182.1549	&	-41.03	&	2.18	&	53303.0139	&	20.51	&	6.85	\\
52180.1209	&	-32.30	&	2.98	&	\multicolumn{3}{c}{\bf\underline{PG2317+046*}}					&	52184.0756	&	-75.94	&	12.40	&	53303.0280	&	10.26	&	7.28	\\
52180.1334	&	-23.46	&	3.35	&	52128.0886	&	-85.62	&	7.24	&	52184.0896	&	-66.75	&	4.12	&	53546.1780	&	15.15	&	3.68	\\
52181.1089	&	-45.07	&	4.79	&	52128.0911	&	-108.52	&	7.53	&	52186.0248	&	-43.18	&	4.25	&	53546.1921	&	19.89	&	4.41	\\
\end{tabular}
  \end{center}
\end{table*}

\end{document}